%% file: LPBIT_ppiv.tex
\pageno=1                                      
\input ppiv-style                                 
\input epsf

\def\wig#1{\mathrel{\hbox{\hbox to 0pt{%
          \lower.5ex\hbox{$\sim$}\hss}\raise.4ex\hbox{$#1$}}}}
\def\mj{M$_{\rm J \;}$}


\def\runningtitletext{Planet Formation}
\def\runningauthortext{Lin et al.}

\null
\firstpageskip

{\baselineskip=14pt
\title{PLANET FORMATION, ORBITAL EVOLUTION}
\title{AND PLANET--STAR TIDAL INTERACTION} 
}

\vskip .3truein
\name{D.N.C. LIN}
\affiliation{UCO/Lick Observatory, University of California, Santa Cruz}
\vskip .2truein
\name{J.C.B. PAPALOIZOU}
\affiliation{Astronomy Unit, Queen Mary \& Westfield College, London}
\vskip .2truein
\name{G. BRYDEN}
\affiliation{UCO/Lick Observatory, University of California, Santa Cruz}
\vskip .2truein
\name{S. IDA}
\affiliation{Dept. of Earth and Planetary Sciences, 
Tokyo Institute of Technology}
\vskip .1truein
\leftline{and}
\vskip .1truein
\name{C. TERQUEM}
\affiliation{UCO/Lick Observatory, University of California, Santa Cruz}
\vskip .3truein

\abstract{
We consider several processes operating during the late stages of
planet formation that can affect observed orbital elements.
Disk-planet interactions, tidal interactions with the central star,
long term orbital instability and the Kozai mechanism are discussed.
}


\mainsection{{I}.~~The formation of protoplanets}

{\it Observations.} It is generally accepted that the planets in our
solar system were formed in a flattened gaseous nebula centered around
the Sun. In typical star--forming molecular clouds, dense cores are
observed to have specific angular momentum larger than $6 \times
10^{20}$~cm$^2$~s$^{-1}$ (Goodman et al. 1993) such that their
collapse leads to rotationally supported disks analogous to the
primordial solar nebula (Terebey et al. 1984). Between 25 to 75\% of
young stellar objects (YSOs) in the Orion nebula appear to have disks
(Prosser et al. 1994; McCaughrean and Stauffer 1994) with typical mass
$M_d \sim 10^{-2 \pm 1}$~M$_\odot$, temperature $\sim 10^{2 \pm 1}$~K
and size $\sim 40 \pm 20$~AU (Beckwith and Sargent 1996; see chapters
by McCaughrean et al, Calvet et al. and Wilner et al.).  The common
existence of protostellar disks around YSOs, with properties similar
to those expected for the solar nebula, suggests that the conditions
for planetary formation may be generally satisfied.

Recent observational breakthroughs have led to the discovery of
Jupiter mass (\mj) planets around at least a few percent of nearby
solar--type stars (see chapter by Cochran et al.).  With the present
data, we can assert that planetary formation is robust.

In {\it conventional planetary formation models} (Pollack et al.
1996), the first stage of protoplanetary formation is the rapid build
up of solid cores through the coagulation of planetesimals (Safronov
1969; Wetherill and Stewart 1989; Lissauer and Stewart 1993; Aarseth
et al. 1993; Kokubu and Ida 1996).  When the core mass increases above
a critical value ($\sim$~a~few~M$_\oplus$), quasi--static evolution is
no longer possible and a rapid accretion phase begins (Mizuno 1980;
Bodenheimer and Pollack 1986) leading to the formation of gaseous
giant planets (Pollack et al. 1996; also see chapter by Wuchterl et
al.).

The details of this model are not yet fully worked out, but if we
suppose the protoplanet can accrete gas as efficiently as possible, it
will first take in gas in the neighborhood of its orbit assumed
circular at radius $r_p,$ until it fills its Roche radius $R_R = (q/3)
^{1/3} r_p$ where $q=M_p/M_*$ is the protoplanet central star mass
ratio, while orbiting in an empty annulus.  The mass of the
protoplanet is then given by $M_p=3 \left[ 4M_d(r)/3M_* \right]^{3/2}
M_*,$ where $M_d(r)=\pi \Sigma r^2$ is the characteristic disk mass
within radius $r,$ with $\Sigma$ being the surface density.  At 5.2~AU
and $\Sigma = 200$~g~cm$^2,$ this gives a mass $M_p \sim 0.4$~M$_{\rm
J}.$

Further mass growth now depends on whether the disk has a kinematic
viscosity $\nu$ capable of producing a mass accretion rate $\dot{M}_p$
onto the protoplanet. If $\nu$ is finite, then all of the mass flow
through the outer disk should go to the protoplanet.  To model the
disk viscosity, we adopt the prescription of Shakura and Sunyaev
(1973) in which $\nu = \alpha H^2 \Omega$, where $\alpha$ is a
dimensionless constant, $H$ is the disk semi-thickness and $\Omega$ is
the disk angular velocity.  The most likely mechanism for providing an
effective viscosity in stellar accretion disks is MHD turbulence
(Balbus and Hawley 1991), which produces $\alpha \wig> 10^{-2}$ in a
fully ionized disk.  Note however that $\alpha$ may vary throughout
the disk and be much smaller in its intermediate parts (see chapter by
Stone et al.).

Observationally inferred values of the disk accretion rate $\dot{M}_d
\sim 10^{-8 \pm 1}$~M$_\odot$~yr$^{-1}$ (Hartmann et al. 1998) are
model dependent and highly uncertain.  Nonetheless, for such a
fiducial $\dot{M}_d$, a protoplanet may attain a mass $M_p > 10$~\mj
within $10^6$~yr. The mass of extra solar planets is $\sim 1$~\mj (see
chapter by Cochran et al.).  Unless these planets are preferentially
formed in low--mass disks, their growth needs to be
terminated/inhibited such that they are unable to accept all the mass
that flows through the disk.  In this paper we focus on
disk/protoplanet tidal interactions as a mechanism for accomplishing
this.

{\it A protoplanet exerts tidal perturbations} which, for $M_p \sim
1$~\mj, may be adequate to induce the formation of a gap in
the disk near its orbit and thereby start to limit accretion flow onto
it (Lin and Papaloizou 1993).  In general, the gravitational potential
due to the protoplanet may be Fourier decomposed in the form:
$$
\psi = \Sigma_{l} \; 
\Sigma_{m=0}^\infty \; \psi_{l,m} {\rm cos} 
[m(\phi-\omega_{l,m} t)] ,
\eqno(1)
$$
where $\phi$ is the azimuthal angle, $l$ and the azimuthal mode number
$m$ are integers and $\omega_{l,m} = \omega + (l-m)
\kappa_p / m$ is the pattern speed, with $\omega$ and $\kappa_p$ being
the angular and epicyclic frequency of the protoplanet,
respectively. If the orbit of the planet has a small eccentricity $e,$
then $\psi_{l,m} \propto e^{\vert l-m \vert}$ (Goldreich and Tremaine
1980; Shu 1984).  For circular orbits, only $l=m$ need be considered.

Both outgoing and ingoing density waves are excited in the disk at the
Lindblad resonances located at $r=r_L$ and where, for $l=m$, $\Omega=
\omega \pm \kappa /m$ (Goldreich and Tremaine 1978).  Here $\kappa$
is the epicyclic frequency of the gas and $\Omega$ is the disk angular
velocity.  In a Keplerian disk, $\kappa = \Omega$.  The
ingoing/outgoing waves carry a negative/positive angular momentum flux
measured in their direction of propagation as they move away from the
protoplanet into the disk interior/exterior.  The waves thus carry a
positive, outward propagating, conserved angular momentum flux or wave
action $F_H.$ In most cases it is reasonable to assume the waves are
dissipated at some location in the disk where their angular momentum
density is deposited (see Lin and Papaloizou 1993 and references
therein).  In the limit of a cold two dimensional disk, Goldreich and
Tremaine (1978) found for a particular $m$ that:
$$
F_H = \left[ { m \pi^2 \Sigma r^2 \over 3 \Omega^2 (m-1)}
\left( {d\psi_{m,m} \over dr} + {2 \Omega \psi_{m,m} \over r 
\left( \Omega - 
\omega_{m,m} \right) }
\right)^2 \right]_{r=r_L} .
\eqno(2)
$$
The back reaction torque exerted on the disk interior (exterior) to
the planet is $-F_H$ ($F_H$). Thus the inner disk loses angular
momentum while the outer disk gains it.  Hence the tendency to form a
gap.  Evaluation of the total torque acting on each side of the disk
requires the summation of contributions from all the resonances which
for circular orbits amounts to summing over $m.$ As $m
\rightarrow \infty,$ the location of the resonance approaches the
orbit.

However, for a non self--gravitating disk and large $m$, the waves are
sonic in character and thus can only exist further than a distance
$\Delta \sim 2 H/3$ from the orbit beyond which the relative disk flow
is supersonic.  This results in a {\it torque cut off} (Goldreich and
Tremaine 1980; Artymowicz 1993a; Korycansky and Pollack 1993; Ward
1997) for $m > r_p/H.$ The dynamics in the coorbital region $\Delta
\wig< 2 H/3$ is not wave--like and is considered below.  Summing over
over all resonances, the total angular momentum flux carried by the
waves is essentially the same as that given by Papaloizou and Lin
(1984), namely:
$$
\dot{H}_T = 0.23 q^2 \Sigma r_p^4 \omega^2 (r_p/H)^3,
\eqno(3)
$$
which was obtained by direct calculation of the torques for a model
disk in which the protoplanet orbited in a gap of width $\sim H.$

Using equation~(3) to estimate tidal torques, and then considering the
competition between viscous torques, which tend to fill a gap, and
tidal torques, which tend to empty it, Lin and Papaloizou (1979a,
1980, 1986a, 1993) found a viscous condition for gap opening: $ q > 40
\nu/ \left( \Omega r_p ^2 \right). $ They proposed that gap formation
would lead to the limitation of accretion but were only able to
consider empty gaps. With the introduction of powerful numerical
finite difference techniques and computers, recently it has become
possible to study gap formation more fully numerically than previously
(Artymowicz et al. 1998; Bryden et al. 1998; Kley 1998; and see
below). The first results of this work indicate that there is a
transitional regime in viscosity for which a gap exists, but some
accretion still occurs, but then being essentially switched off for
small enough viscosity.

In addition, Lin and Papaloizou (1993) pointed out that in a disk with
very small viscosity, in order to form a gap it is necessary that $ H<
R_R = (q/3) ^{1/3} r_p.$ This thermal condition can be viewed in
several ways. It means that the protoplanet's gravity is more
important than pressure at a distance $R_R$ from the protoplanet
(generating large enough hydrostatic pressure forces would require
gradients that would cause a violation of Rayleigh's criterion).  From
Korycansky and Papaloizou (1996) it is apparent that the thermal
condition is required in order that the flow in the neighborhood of
the protoplanet be nonlinear enough that shocks be produced which can
provide the dissipation which is associated with gap opening (see
below). This condition was found for a simple 2D--model disks with a
barotropic equation of state.

It may be possible with more complicated physics and the introduction
of three dimensional effects (Lin et al. 1990a, 1990b) to alter wave
dissipation patterns and thus angular momentum deposition.  However,
because of the difficulty of clearing material near the protoplanet
(it is difficult to see how linear waves could be dissipated closer
than several vertical scale heights), it is likely that gap formation
will occur much less readily if the thermal condition is not
satisfied.

{\it The flow around a partially embedded protoplanet} has been
simulated in the low--$M_p$ limit in a shearing-sheet approximation
(Korycansky and Papaloizou 1996; Miyoshi et al. 1998).  These
simulations provide useful clues on the flow pattern near the
protoplanet.  A global illustrative model is shown in Figure~1, in
which we set $q = 10^{-4}$, $H/r=0.07$, and $\alpha=10^{-3}$.  The
flow pattern can be divided into three regions.  (1) A protoplanetary
disk is formed within $R_R$ of the protoplanet with $H \sim \vert
r-r_p \vert$.  In such a geometrically thick disk, the tidal
perturbations of the host star induce a two-arm spiral shock wave with
an open pitch angle.  Angular momentum is efficiently transfered from
the disk to the star's orbit around the planet (equivalent to the
planet's orbit around the star) such that gas is accreted onto the
core of the protoplanet within few orbital periods $P_{\rm o} = 2\pi /
\omega$.  (2) An extended arc is formed in the coorbital region near
$r_p$ with gas streaming in horseshoe orbits around the L$_4$ and
L$_5$ points.  Because these are local potential maxima, viscous
dissipation leads to the depletion of this region (Lin et al. 1987).
(3) In those regions of the disk which have $\vert r - r_p \vert >
R_R$, the protoplanet's tidal perturbation results in the convergence
of stream lines to form pronounced trailing wakes, both inside and
outside $r_p.$ These high density ridges have been identified by some
(Artymowicz and Lubow 1996) as gas streams along which material flows
from the disk to the protoplanet.  However, the velocity field viewed
in a frame rotating with the companion (Figure~1) clearly indicates
that gas in the post--shock region along the ridge line is moving away
from the protoplanet. 

\vskip .3in

\epsfxsize=\hsize
\epsffile{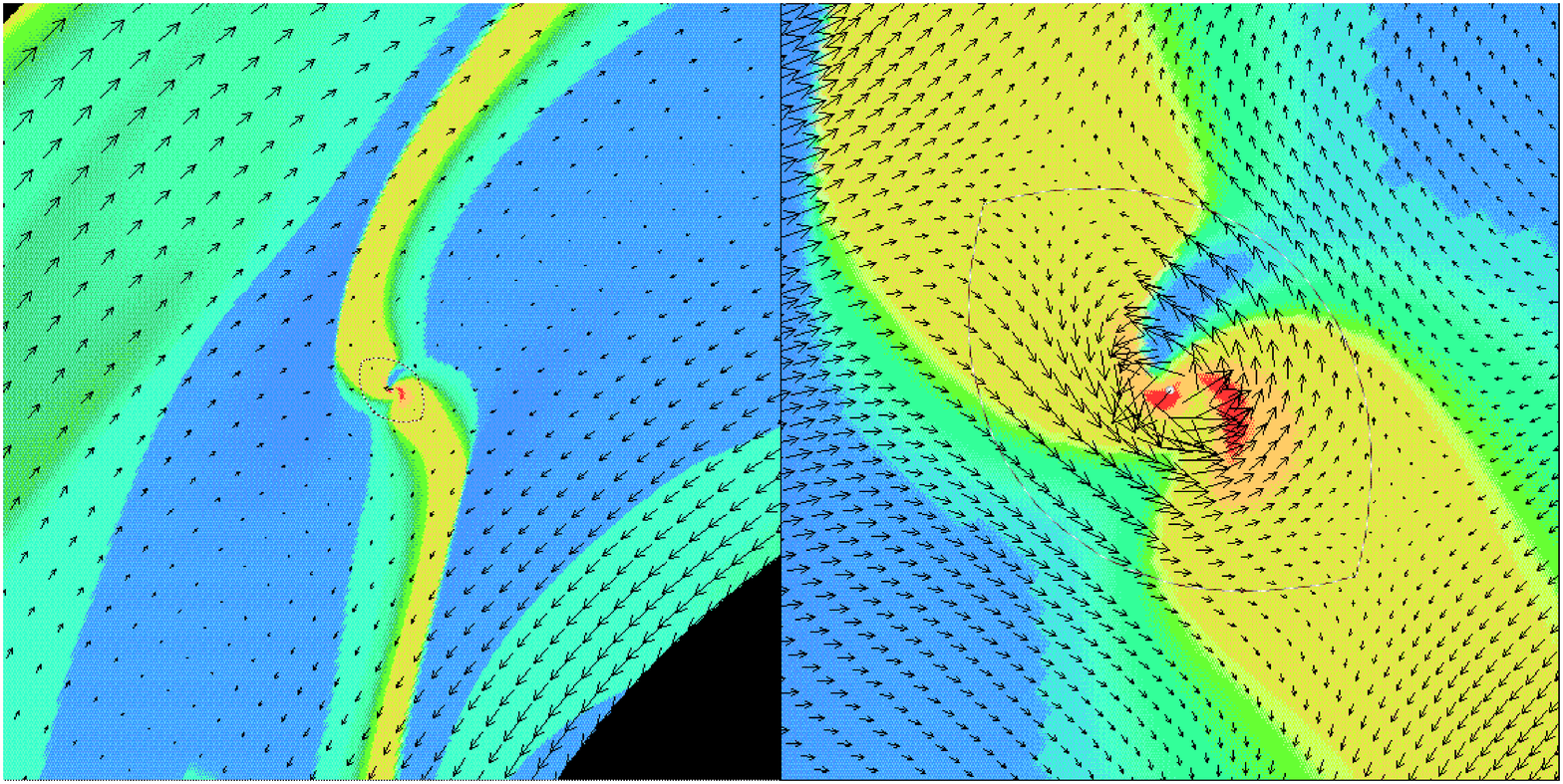}
\vskip .1in
\caption{Figure 1.\capskip The flow pattern around an embedded
protoplanet.  On the left is the global trailing wave (the star is
located toward the lower right).  On the right is a close up view of the
circulatory region immediately surrounding the protoplanet.
Grey scale indicates surface density and
the protoplanet's Roche lobe is shown with a dotted line.
Model parameters are $q=10^{-4}$, $H/r=0.07$, and $\alpha=10^{-3}$.}
\vskip .3in
\tenpoint
\baselineskip=12pt                      

{\it Simulations of gap formation} need to consider model evolution
over many orbital periods and have to deal with a large density
contrast between the gap region and other parts of the disk.  Special
care is needed to minimize the tendency of numerical viscosity to
produce spurious diffusion into the gap region and thus significantly
affect the results.  In Figure~2, we illustrate the excitation and
propagation of waves and the existence of a clear gap for a planet
(with $q=10^{-3}$) interacting with a disk (with $H/r = 0.04 $ and
$\alpha =10^{-3}$).

\vskip .3in

\epsfxsize=\hsize
\epsffile{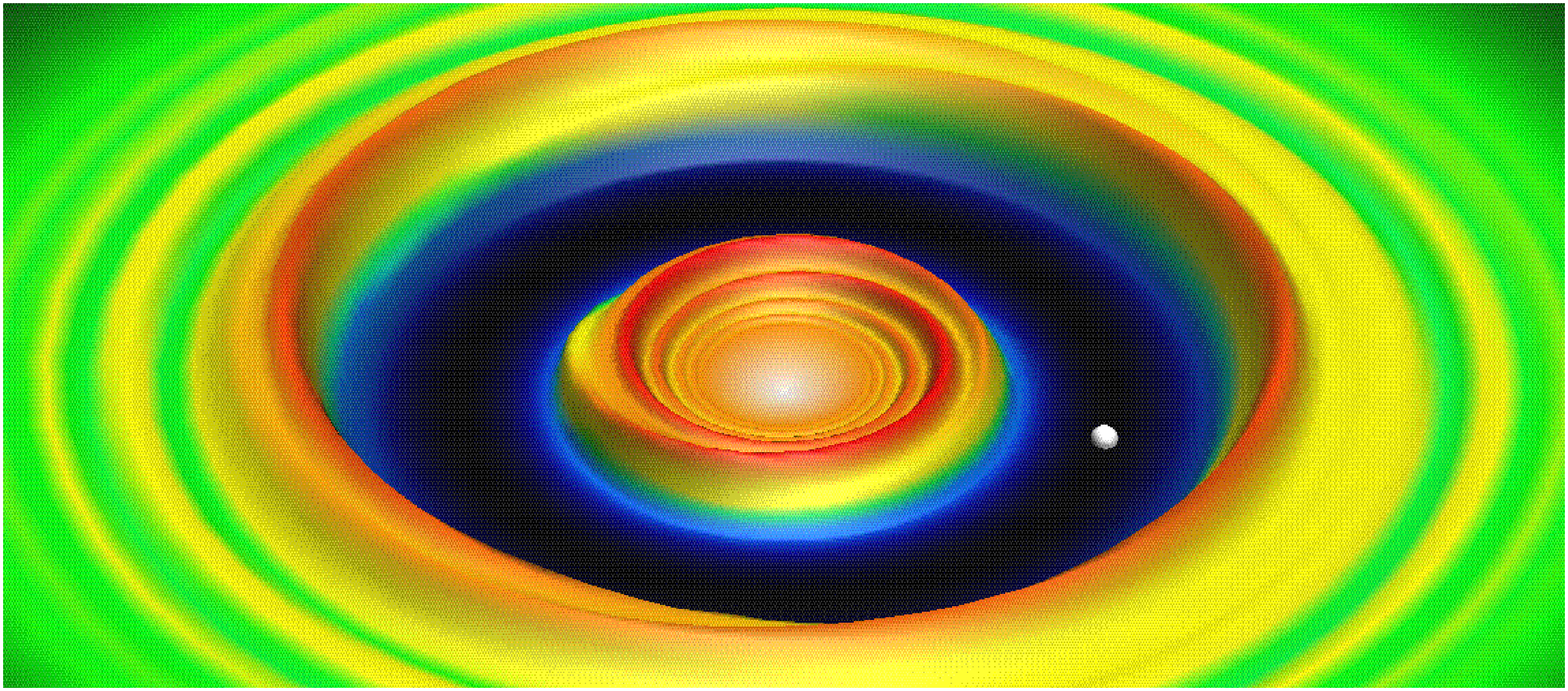}
\vskip .1in
\caption{Figure 2.\capskip Surface density distribution for a gap
opening protoplanet.  The surface density near the planet (indicated
by a white sphere) has been reduced by four orders of magnitude.
Waves are clearly seen propagating both inward and outward away from
the protoplanet. 
Model parameters are $q=10^{-3}$, $H/r=0.04$, and $\alpha=10^{-3}$.}
\vskip .3in

\epsfxsize=\hsize
\epsffile{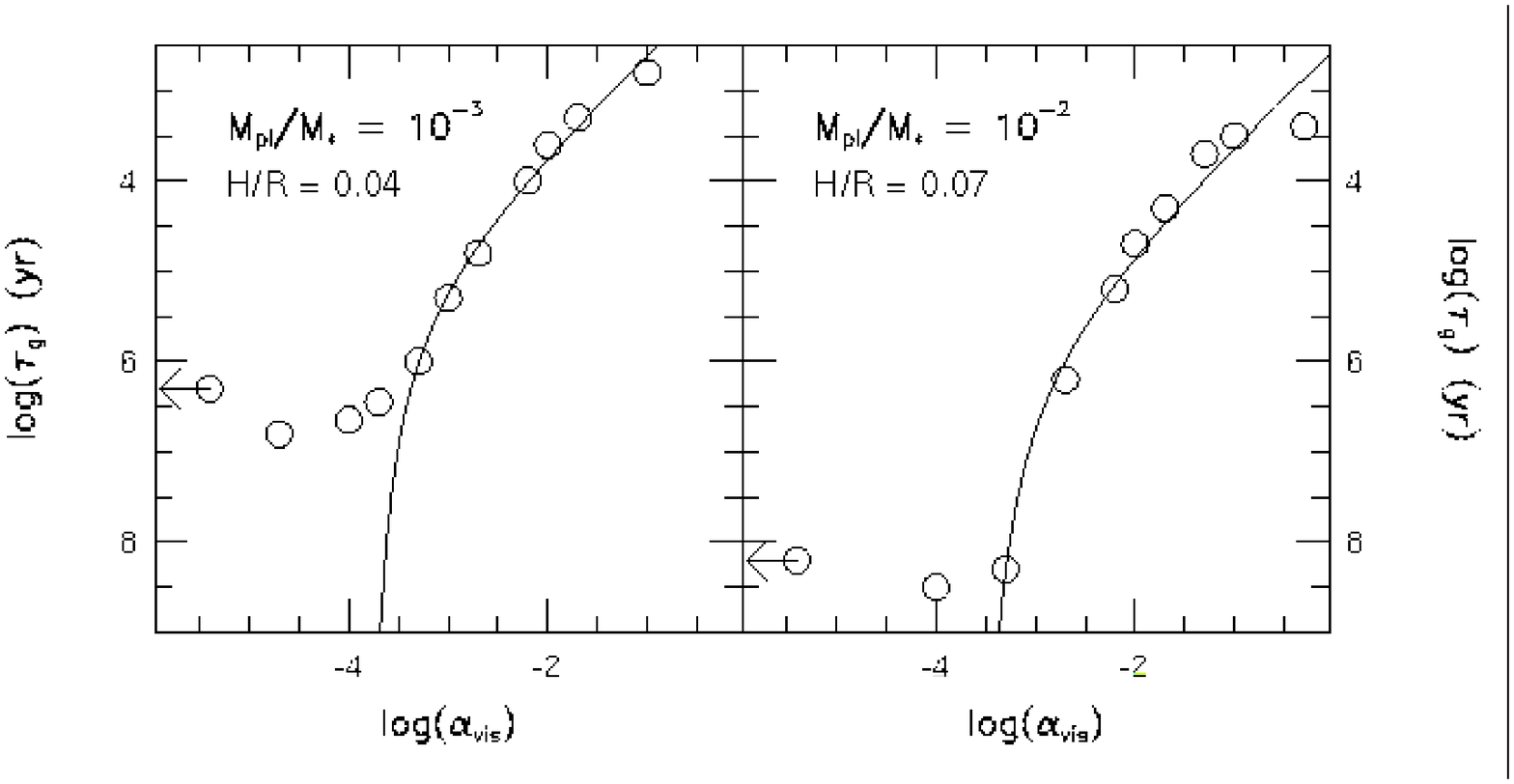}
\vskip 0.1in
\caption{Figure 3.\capskip The mass doubling timescale $\tau_g$ as a
function of viscosity  parameter $\alpha$  for one
Jupiter mass and $H/r=0.04$ (left) and 10 Jupiter masses and
$H/r=0.07$ (right).  
As the gas viscosity is decreased, the accretion  rate onto the protoplanet
drops off sharply ultimately being determined by  numerical effects  for $\alpha < 10^{-4}.$}
\tenpoint
\baselineskip=12pt                      
\vskip .3in

In Figure~3, we plot the growth timescale $M_p / \dot{M}_p$ as a
function of $\alpha$ and $H/r$ for protoplanets of mass 1 and 10~\mj.
We comment that it is not necessary for accretion to stop entirely for
gap formation to affect the final mass of a protoplanet.  All one
needs is that the mass doubling time should become longer than either
the disk lifetime or the time for the interior disk to accrete. In the
latter case, the planet would approach the central star before it
could increase its mass significantly (Ivanov et al. 1998).  From
Figure~3 we see that, for $\alpha \sim 10^{-3}$ and $H/r=0.04$, the
mass doubling time for 1~\mj is $10^{6}$~yr.  For 10~\mj, the same
results apply for $H/r=0.07.$ The results in Figure~3 thus suggest
that the tidal truncation process operates during planetary formation
and is important in determining the final mass of the planet $M_p$.

\mainsection{{I}{I}.~~Protoplanetary orbital evolution}

Resonant interaction between the disk and the planet can lead to
changes in the orbital eccentricity $e$ (Lin and Papaloizou 1993;
Artymowicz 1993b).  The way in which Lindblad torques cause a small
eccentricity to grow has been reviewed by Lin and Papaloizou (1993).
If the Lindblad torques only are considered, the disk matter is able
to tidally pump eccentricity in a similar way to a rapidly rotating
star (see below).  However, corotation resonances in the disk (at
which the pattern speed of a Fourier component of the tidal
perturbation corotates with the disk) also need to be considered (see
above). When these resonances exist, a torque is applied to the disk
at the corotation radius $r_C$ where $\Omega=\omega_{l,m}$ at a rate
given by:
$$
T_{l,m} ^C = - {m \pi^2 \over 2} \left[ \Sigma 
\left( {d \Omega \over dr }
\right)^{-1} {d \varsigma ^{-1} \over dr}
(\psi_{l,m} )^2 \right]_{r=r_C} ,
\eqno(4)
$$
where $\varsigma = \kappa^2 / (2 \Omega \Sigma)$ is the vortensity.
The role of corotation resonances depends strongly on the
$\varsigma$--distribution.  When $\Sigma$ vanishes at some disk edge
or when there is a gap near a low mass perturber, Goldreich and
Tremaine (1980) showed that the action of torques from corotation
resonances results in damping $e$ more effectively than effects from
Lindblad resonances could excite it. Note that this {\it does not}
apply to the situation where a protoplanet is embedded in a disk with
no edges or gap and where tidal interaction is linear.

Assuming $d ln \varsigma / d ln r \sim 1$ in the unperturbed disk,
Artymowicz (1993b) suggested that the eccentricity of low--mass
embedded protoplanets would be damped by tidal interaction.  However,
in the minimum mass solar nebula model where $\Sigma \propto
r^{-3/2}$, vortensity is independent of $r$ and then all corotation
torques vanish (Ward 1993).  Thus, whether low--mass embedded
protoplanets suffer eccentricity damping (Artymowicz 1992) is model
dependent.  For a planet with a sufficiently large mass to open a wide
gap, corotation resonances would be excluded and $e$ would increase on
a timescale $e^2/(d e^2/dt) \sim q^{-1} (M_\ast/M_d) P_{\rm o}$ (Lin
and Papaloizou 1993) which for massive planets can be shorter than the
inferred disk evolution timescale.

If the growth of protoplanets is limited by the formation of a gap of
width $\Delta_g$, tides may cause $e$ to increase until $e \sim
\Delta_g/r_p$ when a potential expansion valid for small $e$ assumed
in the above analysis breaks down.  In a cold disk, the protoplanet's
angular velocity at peri/ap--helion would then be greater/less than
$\Omega (r_p \mp \Delta_g)$, resulting in a torque reversal.  Such a
process could provide a limit to the growth of $e$.  But a
protoplanet's increased radial excursion may also cause the gap to
widen, in which case $e$ might increase to yet larger values.
Analytic treatments and self consistent numerical simulations in the
large $e$ limit have not been carried out so far.

The characterization of extrasolar planets' orbits from forthcoming
observations will provide some useful constraints on the origin of
eccentricity.  The coexistence of several planets with similar $e$ but
very different $M_p$ (or similar $M_p$ but very different $e$) is not
a natural outcome of the excitation of eccentricity by disk tidal
interaction, as this would tend to produce a $(e,M_p)$ relation
(Artymowicz 1992).

{\it Planetary orbital migration} is induced by the difference
between the inner and outer disk torques, $\Delta T$, which react back
on the planet, assumed to remain in a circular orbit, such that:
$$ 
{d r_p \over dt} =  {2  \Delta T \over r_p \Omega M_p}.  
\eqno(5)
$$
Based on recent work indicating that $\Delta T$ is almost always
negative, Ward (1997) suggested that embedded planets migrate inwards
on a timescale $\sim 10^5 (M_p/{\rm M}_\oplus)^{-1}$~yr.  If this
process occurs in protostellar disks, protoplanets would migrate
towards the stellar surface at $r=R_\ast$ once $M_p > 1$~M$_\oplus$
because their growth timescale (Lissauer and Stewart 1993) would then
become longer than this migration timescale.  Resolution of this rapid
migration delemna may require the complete and nonlinear analysis of
the disk response to the protoplanet in the corotation regions which
may be quite complex (see Figures~1 and~2).  However, supposing that
migration occurs and can be stopped near the star (see below), a
scenario leading to the formation of short-period planets has been
proposed (see chapter by Ward and Hahn).

After gap formation, there may still be some residual accretion
depending on the value of $\alpha$ (see above).  Torques are still
exerted between disk and protoplanet and the imbalance due to
differing properties of the disk on either side of the protoplanet
causes orbital migration (Goldreich and Tremaine 1980; Lin and
Papaloizou 1986b; Takeuchi et al. 1996). The estimated effects of
advection of angular momentum through the accretion flow on this
process are usually found to be small.  For small $M_p$, the
protoplanet behaves like a disk particle and so migrates towards the
star. For larger masses, the evolution is slower.  Nonetheless, the
protoplanet is always expected to reach the star before it has time to
double its mass (Ivanov et al. 1998).

The above discussion naturally leads to the suggestion that
short--period planets were formed at several AU away from their host
stars, and subsequently migrated to their present location (Lin et
al. 1996).  However, {\it in situ} formation cannot be completely
ruled out (Bodenheimer 1997).

{\it A protoplanet's migration may be terminated} near $R_\ast$ if (1)
the disk does not extend down to $R_\ast$ or (2) the host star induces
angular momentum transfer to the protoplanet's orbit via tidal effects
(see section III).  For the first possibility, interaction between the
disk and an intense ($>1$~kG) stellar magnetic field has been
suggested (Konigl 1991) as the cause for the modest observed rotation
period ($P_\ast \sim 8$ days) of classical T Tauri stars (Bouvier et
al. 1993; Choi and Herbst 1996).  Such a magnetic field strength may
be consistent with recent measurements (Guenther 1997).  In this
scenario, the stellar field is assumed to induce a cavity out to the
magnetospheric radius ($r_m \sim$ a few $10^{11}$~cm) where $P_\ast$
is equal to the local Keplerian period of the disk.  When a
protoplanet migrates interior to the 2:1 resonance of the gas at
$r_m$, the protoplanet is decoupled from the disk and its migration is
temporarily stalled.  The main uncertainties here are in the observed
distribution of $P_\ast$ (Mathieu et al. 1998).

\mainsection{{I}{I}{I}.~~Planet--star tidal interaction}

As the planet approaches $R_\ast$, the tides raised on the star or
planet by its companion become strong enough so that their dissipation
leads to orbital evolution.  If the distribution of mass of the
perturbed object has the same symmetry as the perturbing potential,
there is no tidal torque resulting from the interaction.  However, in
general, dissipative processes (e.g. radiative damping or turbulent
viscosity) acting on the tides produce a lag between the response of
the star or planet and the perturbing potential, enabling mechanical
energy to be lost and angular momentum to be exchanged between the
rotation of the perturbed object and the orbital motion.  Only if the
system is circular, synchronous and coplanar (i.e. the spin of the
binary components and that of the orbit are parallel) does the tidal
torque vanish (Hut 1980).  However, such an equilibrium state is not
necessarily stable (Counselman~1973; Hut~1980, 1981).

The response of a star/planet to a tidal perturbation is the sum of
two terms, an equilibrium tide and a dynamical tide.  The equilibrium
tide (Darwin 1879) is the shape that the perturbed object would have
if it could adjust instantly to the tidal potential. It is obtained by
balancing the pressure and gravity forces. The dynamical tide contains
the oscillatory response of the star/planet.  It takes into account
the fact that gravity or g~modes can be excited in the convectively
stable layers of the star/planet and that resonances between the tidal
disturbance and the normal modes of the star/planet can occur
(Cowling~1941).

In massive close binaries, which have a convective core and a
radiative envelope, the dynamical tide cannot be neglected because
tidal friction is caused predominantly by the radiative damping of the
tidally--excited modes (Zahn~1975, 1977; Savonije and Papaloizou~1983,
1984, 1997; Papaloizou and Savonije~1985, 1997; Goldreich and Nicholson
1989; Savonije et al. 1995; Kumar et al. 1995). In that case, g~modes,
which are excited mainly near the convective core boundary, propagate
out through the envelope to the atmosphere, carrying energy and
angular momentum. They are damped close to the surface, where the
radiative diffusion time becomes comparable to the forcing period,
thus enabling a net tidal torque to be exerted on the star or planet.

In the case of solar--type binaries, which have a radiative core and a
convective envelope, it was thought until recently that tidal friction
could be well described by the theory of the equilibrium tide, in
which only turbulent dissipation of the equilibrium tide in the
convective envelope is taken into account (Zahn 1977).  However,
recent studies have indicated that this is not the case.  Even if
radiative damping can be ignored, the torque derived using the
equilibrium tide only is 4--6 times larger than that taking into
account the dynamical tide for binary periods of several days (Terquem
et al. 1998a). This is due to the fact that the theory of the
equilibrium tide can in principle be applied only when the
characteristic timescales of the perturbed object are small compared
to the forcing period.  In a solar--type star however, the convective
timescale in the interior region of the convective envelope is as
large as a month.  Furthermore, because of the uncertainty over the
magnitude of the turbulent viscosity associated with convection, it is
not clear that the torque due to turbulent dissipation acting on the
full tide in the convective envelope is more important than that due
to radiative damping acting on the g~modes which propagate inwards in
the radiative core (Terquem et al. 1998a; Goodman and Dickson 1998).
Here, we focus on the case of a planet orbiting around a solar--type
star.

{\it Tides raised on the star by the planet} can be analyzed in the
limit that the rotational frequency of the star is usually small
compared to the orbital frequency.  Then, as a result of tidal
friction, the star spins--up, the orbit decays (the planet spirals in)
and its eccentricity, if any, decreases.  To calculate the timescales
on which this evolution occurs, we need to quantify the dissipative
mechanisms which act on the tides.  Recent studies (Claret and Cunha
1997; Goodman and Oh 1997; Terquem et al. 1998a) have found that the
turbulent viscosity in the convective envelope that is required to
provide the observed circularization rates of main--sequence
solar--type binaries (Mathieu~1994) is at least 50 times larger than
that simply estimated from mixing--length theory for non--rotating
stars.  This indicates either (1) that the observations are
questionable, (2) that solar--type binaries are not circularized
through turbulent viscosity acting on tidal perturbations (see
Tassoul~1988 and Kumar and Goodman 1996 for other suggested tidal
mechanisms), or (3) that dissipation in the convective envelope of
solar--like stars is significantly more efficient than is currently
estimated (see Terquem et al. 1998a for a more detailed discussion).
Here we assume that circularization of solar--type binaries does occur
through the action of turbulent viscosity on the tides, and we then
calibrate its magnitude so as to account for the observed
timescales. Under these circumstances, when the response of the star
is not in resonance with one of its global normal modes, the tides are
dissipated more efficiently by turbulent viscosity than by radiative
damping. In a resonance, radiative damping dominates and limit the
response of the star at its surface.  Although the planetary companion
may go through a succession of resonances as it spirals in under the
action of the tides, for a fixed spectrum of stellar normal modes its
migration is controlled essentially by the non--resonant interaction.
For a non-rotating star, the orbital decay timescale, spin--up
timescale of the star and circularization timescale, in Gyr, are
(Terquem et al. 1998a):
$$
t_{\rm orb}^{\ast} \left( {\rm Gyr} \right) = 2.763 \times 10^{-4} {
\left( M_p / M_{\ast} +1 \right)^{5/3} \over M_p / M_{\ast} }
\left( {P_{\rm o} \over 1 \; {\rm day}} \right)^{13/3} ,
\eqno(6)
$$
$$
t_{\rm sp}^{\ast} \left( {\rm Gyr} \right) = 1.725 \times 10^{-6}
\left( { M_p + M_{\ast} \over M_p } \right)^2
\left( {P_{\rm o} \over 1 \; {\rm day}} \right)^3 ,
\eqno(7)
$$
$$
t_{\rm circ}^{\ast} \left( {\rm Gyr} \right) = 4.605 \times 10^{-5} {
\left( M_p / M_{\ast} +1 \right)^{2/3} \over M_p / M_{\ast} }
\left( {P_{\rm o} \over 1 \; {\rm day}} \right)^{13/3} ,
\eqno(8)
$$
where $P_{\rm o}$ is the orbital period.  This cirularization
timescale is valid only if the initial eccentricity is not too large.
If only the convective envelope of the star, where tidal dissipation
occurs, is spun--up during tidal evolution, then the spin--up
timescale has to be multiplied by $I_c/I$, where $I_c$ and $I$ are the
moments of inertia of the convective envelope and the entire star,
respectively. For the Sun, $I_c/I \simeq 0.14$.

{\it Tides are also raised on the planet by the star.}  Contrary to
the giant planets of our Solar System, Jupiter--like planets on a
close orbit are expected to have an isothermal, and thus radiative
(convectively stable) envelope (Guillot et al. 1996; Saumon et
al. 1996).  For these planets, both turbulent dissipation in the
convective core and radiative damping in the envelope act on the
tides. So far it is not clear which mechanism is more important.

We express the timescales associated with turbulent dissipation of the
tides in terms of the parameter $Q$, the inverse of which is the
effective tidal dissipation function (MacDonald 1964).  Damping of the
tides in the convective core of the planet leads to the
synchronization of the planet rotation with the orbital rotation on
the following characteristic timescale, in Gyr (Goldreich and Soter
1966):
$$
t_{\rm sp}^{\rm p,1} \left( {\rm Gyr} \right) \sim 4.4 \times 10^{-13}
Q \left| {1 \; {\rm day} \over P_p} - {1 \; {\rm day} \over P_{\rm o}}
\right| {M_p \over M_{\ast}}
\left( {P_{\rm o} \over 1 \; {\rm day}} \right)^2
\left( {a \over R_p} \right)^3 ,
\eqno(9)
$$
where $R_p$ is the planet radius, $a$ is the semi--major axis of the
orbit and $P_p$ is the initial value of the planet rotational period
(before it undergoes tidal interaction with the star).  Since the
rotational angular momentum of the planet is in general small compared
to its orbital angular momentum, synchronization of the planet occurs
before any significant orbital evolution can take place.  Once
synchronization is achieved, damping of the tides raised in the planet
always leads to the decay of the orbital eccentricity on a
characteristic timescale which is, in Gyr (Goldreich and Soter 1966):
$$
t_{\rm circ}^{\rm p,1} \left( {\rm Gyr} \right) \sim 2.8 \times
10^{-14} Q {M_p \over M_{\ast}} {P_{\rm o}
\over 1 \; {\rm day}} \left( {a \over R_p} \right)^5 .
\eqno(10)
$$
The tides raised on the planet do not lead to the decay of the orbit
once the planet is synchronized. We note that $Q$ may depend on the
tidal frequency as seen by the planet, and therefore on the rotational
frequency of the planet. Since $t_{\rm circ}^{\rm p,1}$ is calculated
assuming synchronization, the value of $Q$ in the above equation may
be different from that used for calculating $t_{\rm sp}^{\rm p,1}$.
Orbital evolution of Jupiter's satellites leads to an estimate of $Q
\sim 10^5-10^6$ for this planet (Goldreich and Soter 1966; Lin and
Papaloizou 1979b).  However, it is not yet understood where this value
of $Q$ comes from, since turbulent viscosity arising from convection
would produce $Q \simeq 5 \times 10^{13}$ (Goldreich and Nicholson
1977; see Stevenson 1983 for an alternative).  There is of course no
reason to assume that the $Q$--value of Jupiter is similar to that of
the extra--solar planets.  First, some of these planets orbit very
close to their parent star, and therefore are much hotter than
Jupiter. Also, $Q$ may depend on the magnitude and the frequency of
the tidal oscillation, in which case it would be different for Jupiter
if this planet were synchronously rotating on a closer orbit.
Therefore we can only speculate when applying the above formulae to
extra--solar planets.

Radiative damping of the tides in the envelope, as described above for
massive binaries, gives rise to the synchronization and
circularization timescales $t_{\rm sp}^{\rm p,2}$ and $t_{\rm
circ}^{\rm p,2}$, respectively.  These timescales have been evaluated
(in particular for 51~peg) by Lubow et al. (1997).  However, as they
pointed out, the asymptotic analysis they use is valid only if the
initial spin rate of the planet is smaller than half that of Jupiter.
Besides, this analysis neglects the effect of rotation on the tides,
which is important for near--synchronous planets.

{\it In the context of extrasolar planets,} we first comment on the
magnitude of the perturbed velocity induced by the tides at the
stellar surface.  Terquem et al. (1998a, 1998b) have found that in the
case of 51~Peg, this velocity is too small to be observed. This result
is insensitive to the magnitude of the stellar turbulent viscosity and
is not affected by the possibility of resonance. It also holds for the
other extra--solar planets which have been detected so far.

The discussion of the previous subsection indicates that the rotation
of planets on a close orbit is almost certainly synchronous with the
orbital rotation.  For the star to be synchronized in less than 5~Gyr,
i.e. $t_{\rm sp}^{\ast} < 5$~Gyr, a 1 (3) Jupiter mass planet would
have to be on an orbit with a period less than 1.4 (3) day(s).  We
note that in the case of $\tau$~Boo, the rotation of the star may then
have been synchronized as a result of tidal effects.  If only the
convective envelope of the star is spun--up, these periods have to be
multiplied by about a factor 2. Therefore, the observation of
anomalous rapid rotators could give further evidence of the presence
of close planets and provide some indication on how the dissipation of
tides affect the rotation of the star (Marcy et al. 1997).

The circularization timescale for the orbit is $t_{\rm circ}$ such
that $1/t_{\rm circ} = 1/t_{\rm circ}^{\ast} + 1/t_{\rm circ}^{\rm
p,1} + 1/t_{\rm circ}^{\rm p,2}$.  We note that, according to the
expression of the timescales we have given above, $t_{\rm orb}^{\ast}
\simeq 6t_{\rm circ}^{\ast}$ and $t_{\rm orb}^{\rm p,1} > 10 t_{\rm
circ}^{\rm p,1}$, so that circularization can be achieved without a
significant orbital decay taking place (in contrast to the statement
by Rasio et al. 1996).  So far it is not clear which of the timescales
$t_{\rm circ}^{\ast}$, $t_{\rm circ}^{\rm p,1}$ and $t_{\rm circ}^{\rm
p,2}$ is the shortest.  For 51~Peg, $t_{\rm circ}^{\rm p,1} < t_{\rm
circ}^{\ast}$ requires $Q<2 \times 10^7$ (see also Rasio et al. 1996
and Marcy et al. 1997 for estimates of $t_{\rm circ}^{\rm p,1}$).  We
point out that for a Jupiter mass planet, $t_{\rm circ}^{\ast} <
5$~Gyr for $P_{\rm o} < 3$~days. Therefore, {\it all the Jupiter mass
planets orbiting with a period less than about 3~days should be on a
circular orbit}.  This cutoff period is a lower limit, since $t_{\rm
circ}^{\ast}$ is an upper limit of the circularization timescale.

Planets found on eccentric orbits at smaller periods are indicative
that the stellar rotation frequency (assumed zero up to now) is large
compared to the orbital frequency.  Indeed, in that case tidal
friction in the star pumps up the orbital eccentricity, opposing the
effect of tidal friction in the planet.  However, the above
calculations indicate that the planet may have to be on a very close
orbit for the tidal friction in the star to be able to increase
significantly the orbital eccentricity.  Whether the planet can get to
such a close orbit is questionable if the disk in which the planet
formed did not extend down to $R_{\ast}$.

The timescale on which the orbit decays is $t_{\rm orb}^{\ast}$. The
condition $t_{\rm orb}^{\ast} < 5$~Gyr requires $P_{\rm o}<2$~days for
a Jupiter mass planet.  If a planet has not been able to get to such a
close orbit because, e.g., the inner parts of the disk were truncated,
orbital decay may not occur as a result of tidal friction, and it may
not plunge onto their parent star.

However, {\it several planets may attain short periods}.  After the
first-born protoplanet emerges and migrates to a few $R_\ast$, because
the formation timescale increases with $r,$ additional, subsequently
formed protoplanets may migrate inwards pushing protoplanetary cores
ahead of their inward path until they become trapped at Lindblad
resonances.  Such resonant trapping is found for the Galilean
satellites (e.g. Goldreich and Peale 1966; Lin and Papaloizou 1979b)
and it occurs if the migration rate is sufficiently slow.  It raises
the intriguing possibility that planets temporarily parked close to
the star because of absence of disk material could be forced to
plunge into the host star.

On the other hand, if the star is a rapid rotator (with period shorter
than that of the orbit) and has a weak magnetic field, tidal effects
transfer angular momentum from the star to the orbit with an
increasing efficiency as the orbit is pushed out until a tidal barrier
is produced such that the migration time and inverse orbital decay
time balance.  Tentatively using estimates derived from equation~(7),
an inward migration rate of $10^6$~yr could only be balanced by tidal
effects acting on a 1~\mj planet if the period was $\sim 0.3$~days,
which increases to $\sim 0.7$~days for a 3~\mj planet and a migration
time of $10^7$~yr.  If planets can survive at such short periods (more
likely for massive planets and slow migration rates) and if
eccentricity growth is suppressed by tides within the planet, tidal
effects might temporarily cause the halting of migration by
transferring angular momentum from the star to the inner orbit and from
there to any residual disk material via Lindblad torques and so on to
outer planets (cf. the situation for Saturn's rings).  Such processes
have not yet been worked out in detail.  However, a solar--type star
is likely to have a small amount of angular momentum in comparison to
the disk/planet system in total, and once the star slows down
sufficiently, the inward migration must continue.

As a planet eventually becomes engulfed by the host star, it is
disrupted by tidal breakup, heating, and ram pressure stripping
(Sand\-quist et al. 1998).  Supporting evidence for such phenomena may
be found in the super--solar metal abundance of 51~Peg (G2V),
55~$\rho$~Cnc (G8V), and $\tau$~Boo (F7V) (Butler et al. 1997;
Gonzalez 1997, 1998).  The convective envelope for each of these
stars contains a few $10^{-2}$~M$_\odot$.  The mixing of 10--40
M$_\oplus$ of planetary heavy element material within the stellar
envelope (Zharkov and Gudkova 1991) would lead to a significant metal
enrichment there.  Since the depth of the convection zone of main
sequence stars decreases significantly with increasing stellar mass,
the routine engulfing of planets by their hosts might lead to a
tendency for hotter planetary hosts to show a general overall
metallicity enhancement with respect to cooler ones.

\mainsection{{I}{V}.~~Long term stability of planetary systems}

In a relatively massive disk, several giant planets may be formed with
$M_p \sim$~1--3~\mj and $a > 1$~AU. Their long term orbital stability
determines the dynamical evolution of the system.  After the depletion
of the disk gas, mutual gravitational perturbation between the planets
 may gradually increase their eccentricities until their
orbits cross each other on a timescale $\tau_x$.  Extrapolation of
existing numerical results (e.g. Franklin et al. 1990; Chambers et
al. 1996) give $\tau_x \sim 10^{12-18}$~yr for the solar system.  But
$\tau_x$ would reduce to $\sim 10^{8-12}$ or $10^{5-8}$~yr if all the
gas giants had $M_p=$1 or 2~\mj, respectively.  

Since systems of massive planets formed with similar values of $a$ and
small $e$ eventually suffer orbit crossing (Lin and Ida 1997), massive
eccentric planets may have acquired their orbital properties as a
consequence of orbit crossing (Rasio and Ford 1996; Weidenschilling
and Marzari 1996; Lin and Ida 1997).  Once orbit crossing occurs,
close encounters eventually take place. When these involve equal
mass planets, they produce a velocity perturbation $V$ with magnitude
limited by the surface escape velocity $V_{\rm esc}
\simeq 60 \left( M_p / {\rm M}_{\rm J}
\right)^{1/3} \left( \rho_p / 1 \; {\rm g \; cm}^{-3} \right)^{1/6}$ 
km~s$^{-1}$, where $\rho_p$ is the pla\-net's internal density
(Safronov 1969).  This results in $e \wig< V_{\rm esc}/(a
\Omega) \sim 2 \left( M_p/ {\rm M}_{\rm J} \right)^{1/3}
\left( a / 1 \; {\rm AU} \right)^{1/2}$.
Thus observed eccentricities can be accounted for.

Supposing that giant planets form at $\sim 1$~AU, massive planets
($M_p \wig> 5$~\mj) with moderately high $e \sim 0.3$ and moderately
small $a \sim 0.5$~AU can be accounted for by orbit crossing and
merging.  Eccentric planets with $a > 1$~AU and $M_p
\sim$~\mj can be produced by orbit crossing followed by ejection.
Merger, resulting in $e \wig< 0.5$, is favored at small $r$ whereas
ejection occurs preferentially at large $r$ because the cross--section
for direct collisions is independent of $r$ while that for close
encounters is $\propto V^{-4} \propto (r \Omega) ^{-4} \propto r^2$
(Lin and Ida 1997).  Ida and Lin (1998) find that the orbital
properties of a merged body are consistent with those of the planets
in eccentric orbits around 70~Vir and HD~114762.

Numerical simulations of many planet systems indicate that although
some may be ejected, a residual population of eccentric planets ($e
\wig> 0.5$) may remain bound to the central star at large distances
($a >$ 100~AU).  Close encounters also excite the relative inclination
up to $\sim e/2$~radians.  Detection of additional companions around
stars with an eccentric planet would provide tests of these scenarios.

Rasio and Ford (1996) suggested that the short period planets were
dynamically scattered into a region close to the central star.
Although the short-period planets may be circularized (see above),
numerical simulations indicate that the scattering scenario would lead
to a population of planets with high eccentricity at large distances
(10--100~AU) from their central stars.  A comprehensive search for
these planets would provide a useful test of this scenario as against
that of disk--planet interaction which produces orbits with smaller
$e.$.

\mainsection{{V}.~~Effects of secular perturbations due to a 
distant companion}

Some extrasolar planets are found in binary systems.  For example, a
planet is found around 16~Cyg~B which orbits at a distance of 1.7~AU
from the central star which is is known to have a binary companion
16~Cyg~A.  It has been suggested that the high $e$ (=0.67) of 16~Cyg~B
is excited by the gyroscopic perturbation due to 16~Cyg~A (Holman et
al. 1997).  A similar effect may be caused by the gravitational
perturbations from other planetary bodies.  Here we discuss the effect
of secular perturbations on a long timescale such that a time average
may be performed.  The orbiting bodies may then be considered as
having their mass distributed continuously around their orbits as in
the classical theories of Laplace and Lagrange (see Hagihara 1972 and
references therein).  Semi--major axes do not change under secular
perturbation.  However, changes to the eccentricity and inclination
may be produced.

If orbits are coplanar, in general only modest changes to
eccentricities are produced if the perturbing bodies are widely
separated.  An exception to this might occur if secular resonances
sweep through the system because of a changing gravitational potential
due to disk dispersal or changing stellar oblateness (eg. Ward 1981).
However, if orbits are allowed to have high inclination, large
eccentricity changes can occur to the orbit of an inner planet
perturbed by a distant body or binary companion.

To consider this {\it gyroscopic perturbation effect}, let us consider
the simplest case of the motion of an inner planet with mass $M_p$
perturbed by an outer companion with mass $M'$ assumed to be in a
circular orbit with semi-major axis $a'.$ The outer companion is
assumed to contain sufficient angular momentum that its orbit remains
fixed defining a reference plane to which the inner planet orbit has
inclination $i.$ We note $\omega_a$ the longitude of the apsidal line
measured in the orbital plane from the line of nodes.

If the distance of $M_p$ from $M_*$ is $r,$ its speed is $v,$ the
angle between its position vector and that of $M'$ is $\Theta,$ and
only the dominant quadrupole term in the interaction potential
multipole expansion is retained, its motion is governed by the
Hamiltonian (Kozai 1962; Hagihara 1972; Innanen et. al. 1997):
$$
{\rm H}= {1\over 2} M_* v^2 -{ {GM_* M_p} \over {r} } -{ {GM'M_p r^2}
\over {2a'^3} } \left( 3 \cos^2 \Theta - 1 \right).
\eqno(11)
$$
After performing a time average, appropriate for secular
perturbations:
$$
{\rm H}= -{{GM_* M_p}\over{2a}} -{{GM'M_p a^2}\over{16a'^3}} 
\left[ \left( 3\cos^2i-1\right)(2+3e^2) \right.
$$
$$
\; \; \; \; \; \; \; \; \; \; \; \; \; \; \; \; \; \; \; \; \; 
\; \; \; \; \; \; \; \; \; \; \; \; \; \; \; \; \; \; \; \; \; 
\; \; \; \; \; \; \; \; \; \; \; \; \; \; \; \; \; \; \; \; \; 
\; \; \; \;
\left. + 15e^2\sin^2i\cos2\omega_a \right] .
\eqno(12)
$$
The fact that both H and the component of angular momentum parallel to
the outer orbital axis, $ {\rm L} \cos i = \sqrt{GM_* a(1-e^2)}
\cos i,$ are constant enables elimination of $i$ and $\omega_a,$ so that
a complete solution can be found from the equation for the rate of
change of $e$, which can be derived from the canonical equation $d{\rm
L}/dt = - \partial {\rm H}/\partial
\omega_a,$ in the form:
$$
{{de}\over{dt}} ={{15 n M' a^3 e \sin^2i \; \sin2\omega_a}\over{8 M_* a'^3}},
\eqno(13)
$$
with $2\pi /n$ being the orbital period of the planet $M_p.$

The above equation indicates that $e$ can oscillate between extremes
occurring when $\sin2\omega_a =0.$ There has been particular interest
in finding conditions under which $e$ can start from very small values
and then increase unstably to values $\wig< 1$.  To examine conditions
for this to occur, we use H=const and L$\cos i$=const to find
$\omega_a$ and $i$ in terms of $e.$ Assuming initial values $i=i_0$
and $e=e_0,$ when $\omega_a=\omega_0$, for infinitesimally small but
non zero values of $e_0$ that can be neglected, we find:
$$
\cos2\omega_a = {{1- 5\cos^2i_0 -e^2}\over { 5(1-e^2-\cos^2 i_0)}} .
\eqno(14)
$$
From the above, we see that when $e \rightarrow 0$: 
$$
\cos2\omega_a = {{(1- 5\cos^2i_0)}\over { 5(1-\cos^2 i_0)}}.
\eqno(15)
$$
Clearly, for solutions of the type we seek, we require 
$cos2\omega_a > -1,$ which requires that the initial inclination
exceed a critical value (see Innanen et. al. 1997) such that:
$$
\cos^2 i_0 < 3/5. 
\eqno(16)
$$
If this inequality is satisfied (if it is not then $e_0$ cannot be
neglected), equation~(13) indicates that $e$ grows from a very small
value up until the value obtained by setting $\omega_a= \pi /2,$
namely:
$$
e^2 = 1 - 5 \cos^2 i_0 / 3 ,
\eqno(17)
$$
attained when $\cos^2 i = 3/5.$ A range of eccentricities may be
generated in this way. However, values close to unity may be attained
for initial inclinations close to 90$^0$. There are then solutions
which have large amplitude oscillations in eccentricity (Kozai effect)
and adding back in the non secular terms can lead to chaotic
behavior.  The large eccentricity changes  occur independently of
the size of the perturbation. However,
the characteristic timescale for the eccentricity changes to occur is
given from equation~(13) as $\sim ( M_* a'^3)/ (M' a^3) $ orbital
periods of the inner planet, which is long for small perturbations.

But note that the effect requires the motion of the apsidal angle
$\omega_a$ be governed only by the perturbation considered. Other
effects may disrupt this, such as general relativistic corrections
(e.g. Holman et al. 1997) and the oblateness of the central star.
Both could be considered in the discussion below, but here we limit
ourselves to considering the effects of relativistic apsidal
precession which may lead to constraints on the orbital elements of
planets with short periods.

To incorporate {\it relativistic apsidal precession} to lowest order,
we modify the Newtonian potential such that keplerian elliptic orbits
are induced to precess at the correct rate.  We modify the potential
due to the central mass, such that:
$$
{{-GM_*}\over {r}} \rightarrow {{-GM_*}\over {r}}
\left(1+{3{GM_*}\over {r c^2}}\right),
\eqno(18)
$$
with $c$ being the speed of light.  The additional term can be added
into the Hamiltonian H and averaged so that equation~(12) becomes:
$$
{\rm H} \rightarrow {\rm H} -{{3(GM_*)^2 M_p}\over{2a^2 c^2
{\sqrt(1-e^2)} }}.
\eqno(19)
$$
The same procedure as described above can be used to find $\omega_a$
and $i$ in terms of $e$ and hence conditions that must be satisfied in
order that large values of $e$ might be attained.
The condition analogous to (16) for the Kozai effect to work is:
$$
\cos^2 i_0 < (3-2f)/5.
\eqno(20)
$$
Here the parameter:
$$
f= \left({{GM_*}\over {a  c^2}}\right)\left( {{M_*a'^3}\over
{M'a^3}}\right)
\eqno(21)
$$
measures the importance of relativistic precession relative to the
perturbation due to the outer companion.  From this, if $f>3/2$ the
effect is completely suppressed.  When the Kozai effect occurs, the
maximum possible eccentricity that can be generated is obtained by
setting $\omega_a= \pi/2$ and $i_0= 90^0$. This gives:
$$
e^2  =  1/2 + \sqrt{(4f/3+1/4}) -4f/3 ,
\eqno(22)
$$
which indicates that, for $f=0.1,$ $e$ as high as 0.99 can be
generated for $i_0 =90^0.$ The parameter $f$ may be conveniently
expressed as:
$$ 
f= 5.0519 \times 10^{-7} {1\over q'} 
\left( {P'_{\rm o} \over P_{\rm o}} \right)^2 
\left( {1 \; {\rm day} \over P_{\rm o}} \right)^{2/3}.
\eqno(23)
$$ 
Here, $P_{\rm o}$ and $P'_{\rm o}$ denote the orbital periods of $M_p$
and $M'$ respectively and $q'=M'/M_*.$

Planets with periods less than around 3--4 days are likely to have
their orbits circularized as a result of tidal effects (see
above). From the requirement $f>3/2,$ we see that for inner planets
with somewhat larger periods, but still on the order of days, binary
companions with mass ratio of order unity need orbital periods shorter
than about $40$~yr in order to pump significant eccentricity.  If a
massive planet with $q' \sim 10^{-2}$ is considered, it must orbit
with a period ten times shorter $\sim 4$~yr. Such an object should be
readily detectable.

We comment that for a given distant binary companion with inclination
close to $90^0$ to the planet orbit, and no other perturbers, there is
a planet with a period long enough such that the Kozai effect works
such that it could produce eccentricities close enough to unity that
the planet has arbitrarily close approaches to the central star. Then
tidal effects may act towards circularization so decreasing the
distance to apoapse until a circular orbit is produced. The final
period would have to be a few days.  Alternatively, should the closest
approach to the star be beyond the largest radius at which orbits can
be circularized, the orbit would retain a Kozai cycle with high
maximum eccentricity.

\mainsection{{V}{I}.~~Summary}

Here we summarize the various processes discussed in the paper, their
influence on the evolution of planetary systems, and their
observational implications.

\noindent

a) {\it Planet-disk tidal interaction} excites potentially observable
spiral density waves and can create clean gaps in the disk, limits
mass growth and drives orbital migration.  Growth limitation by gap
formation together with inward migration may naturally account for the
upper limit in the observed mass ratio distribution of extra solar
planets provided they were formed in disks with $H \wig<0.1 r_p$ and
$\alpha \wig< 10^{-3}.$ Since the critical mass for gap formation is
an order of magnitude larger than for dynamical gas accretion, a
bimodal mass distribution with a depression between these masses may
be anticipated.  After gap formation, massive protoplanets are
expected to migrate with the viscous evolution of the disk until they
encounter a stellar magnetospheric cavity or tidal barrier.  The
existence of short period planets in CTTS would imply that the local
disk viscous evolution timescale is shorter than the typical age
($\sim 10^6$ yr.)  Orbital migration may lead to stellar consumption
of protoplanets with surviving planets locked into commensurable
orbits. Contamination of their host stars (during the latter's main
sequence evolution) by plunging planets may have led to a relatively
high metallicity.  Planet-disk interaction is likely to excite only
small $e \wig < 0.3$ and lead to a relation between $M_p$ and $e$
which could be observed.

{\it Stellar tides raised on short-period planets} should rapidly
synchronizes their spin.  Short period planets with $M_p \sim$~\mj and
$P \wig< 3-4 $~days (e.g. $\tau$ Boo) should have their orbits
circularized by tides raised on a slowly rotating star within a few
Gyr.  In the case of massive planets, the stellar rotation may also be
synchronized with the orbit.  But, only very short period planets ($P
\wig< 2 $~days) are expected to undergo significant orbital decay.

c) {\it Long term gravitational interaction between planets with
initially small $e$} leads to orbit crossing on a time scale which is
sensitively determined by their masses and initial separations.
Subsequent close encounters between comparable masses can lead to high
orbital eccentricity and planets scattered into extended orbits.  This
mechanism may provide a supply of planets with a range of short
periods, some of which may undergo tidal circularization and/or plunge
into the star, together with outlying accomplices. The latter could be
imaged by the next generation of IR interferometers.

d) {\it Secular perturbations} by distant binary, or massive outlying
planetary, companions in relatively highly inclined orbits can also
excite high $e$ through the Kozai mechanism.  Potentially this process
might produce a supply of short period planets ($P \wig< 2-3 $~days)
that have undergone tidal circularization. But because of the very
high relative inclinations required, it is likely to work only in a
small number of cases.
 
This work is supported by NSF and NASA through grants AST-9315578 and
NAGW-4967.

\vfill\eject
\null


\vskip .5in
\centerline{\bf REFERENCES}
\vskip .25in

\ref{Aarseth, S. J., Lin, D. N. C., and Palmer, P. L. 1993.
Evolution of Planetesimals. II. Numerical Simulations.
{\refit Astrophys.\ J.\/} 403:351--372.}

\ref{Artymowicz, P.  1992.
Dynamics of binary and planetary-system interaction with disks -
Eccentricity changes.  {\refit Publ.\ Astron.\ Soc.\ Pacific\/}
104:769--774.}

\ref{Artymowicz, P. 1993a. On the Wave Excitation and a Generalized 
Torque Formula for Lindblad Resonances Excited by External Potential.
{\refit Astrophys.\ J.\/} 419:155--165.}

\ref{Artymowicz, P.  1993b.  Disk-Satellite Interaction via Density 
Waves and the Eccentricity Evolution of Bodies Embedded in Disks.
{\refit Astrophys.\ J.\/} 419:166--180.}

\ref{Artymowicz, P., and  Lubow, S. 1996.
Mass Flow through Gaps in Circumbinary Disks.  {\refit Astrophys.\
J.\/} 467:L77--L80.}

\ref{Artymowicz, P., Lubow, S., and  Kley, W. 1998. 
In {\refit Planetary Systems: The Long View \/}, ed. \ L. Celnikier
(Editions Fronti\`eres), {\refit in press}}

\ref{Bryden, G., Chen, X., Lin, D. N. C., Nelson, R., 
and Papaloizou, J. C. B. 1998.  Tidally induced gap formation in
protostellar disks: gap clearing and suppression of protoplanetary
growth.  {\refit Astrophys.\ J.\/} {\refit Submitted}.}

\ref{Balbus, S. A., and Hawley, J. F. 1991.
A powerful local shear instability in weakly magnetized disks. I ---
Linear analysis. {\refit Astrophys.\ J.\/} 376:214--233.}

\ref{Beckwith, S. V. W., and Sargent, A. I. 1996.
Circumstellar disks and the search for neighbouring planetary systems.
{\refit Nature} 383:139--144.}

\ref{Bodenheimer, P. and Pollack, J. B. 1986.
Calculations of the accretion and evolution of giant planets: The
effects of solid cores.  {\refit Icarus} 67:391--408.}

\ref{Bouvier, J., Cabrit, S., Fernandez, M., Martin, E. L., 
Matthews, J. M. 1993.  COYOTES-I: the photometric variability and
rotational evolution of T-Tauri stars.  {\refit Astron.\ Astrophys.\/}
272:176--206.}

\ref{Butler, R. P., Marcy, G. W., Williams, E., Hauser, H., and 
Shirts, P. 1997.  Three New "51 Pegasi--Type" Planets.  {\refit
Astrophys.\ J.\/} 474:L115--L118.}

\ref{Chambers, J. E., Wetherill, G. W., and Boss, A. P. 1996.
The Stability of Multi-Planet Systems.  {\refit Icarus} 119:261--268.}

\ref{Choi, P. I., and Herbst, W. 1996. 
Rotation periods of stars in the Orion nebula cluster: the bimodal
distribution.  {\refit Astron.\ J.\/} 111:283--298.}

\ref{Claret, A., and Cunha, N. C. S. 1997. Circularization and 
synchronization times in main--sequence of detached eclipsing binaries
--- II. Using the formalisms by Zahn.  {\refit Astron.\ Astrophys.\/}
318:187--197.}

\ref{Counselman, C. C. III 1973. Outcomes of tidal evolution. 
{\refit Astrophys.\ J.\/} 180:307--316.}

\ref{Cowling, T. G. 1941. The non--radial oscillations of polytropic 
stars.  {\refit Mon.\ Not.\ Roy.\ Astron.\ Soc.\/} 101:367--375.}

\ref{Darwin, G. H. 1879.  
{\refit Phil.\ Trans.\ Roy.\ Soc.\ London} 170:1.}

\ref{Franklin, F., Lecar, M., and Quinn, T. 1990.
On the stability of orbits in the solar system - A comparison of a
mapping with a numerical integration.  {\refit Icarus} 88:97--103.}

\ref{Goldreich, P., and Nicholson, P. D. 1977. Turbulent viscosity 
and Jupiter's tidal Q. {\refit Icarus} 30:301--304.}

\ref{Goldreich, P., and Nicholson, P. D. 1989. Tidal friction in 
early--type stars.  {\refit Astrophys.\ J.\/} 342:1079--1084.}

\ref{Goldreich, P., and Peale, S. 1966. 
Spin-orbit coupling in the solar system.  {\refit Astron.\ J.\/}
71:425--438.}

\ref{Goldreich, P., and Soter, S. 1966. $Q$ in the solar system. 
{\refit Icarus} 5:375--389.}

\ref{Goldreich, P., and  Tremaine, S. 1978.
The velocity dispersion in Saturn's rings.  {\refit Icarus}
34:227--239.}

\ref{Goldreich, P., and  Tremaine, S. 1980.
Disk-satellite interactions.  {\refit Astrophys.\ J.\/} 241:425--441.}

\ref{Gonzalez, G. 1997.
The stellar metallicity-giant planet connection.  {\refit Mon.\ Not.\
Roy.\ Astron.\ Soc.\/} 285:403--412.}

\ref{Gonzalez, G. 1998.
Spectroscopic analyses of the parent stars of extrasolar planetary
system candidates.  {\refit Astron.\ Astrophys.\/} 334:221--238.}

\ref{Goodman, A. A., Benson, P. J., Fuller, G. A., and Myers, 
P. C. 1993.  Dense cores in dark clouds. VIII --- Velocity
gradients. {\refit Astrophys.\ J.\/} 406:528--547.}

\ref{Goodman, J., and Dickson, E. S. 1998. 
Dynamical tide in solar-type binaries. {\refit Astrophys.\ J.\/}
{\refit Submitted}.}

\ref{Goodman, J., and Oh, S. P. 1997. Fast tides in slow stars: the 
efficiency of eddy viscosity.  {\refit Astrophys.\ J.\/}
486:403--412.}

\ref{Guenther, E. W. 1997.
Magnetic fields of T Tauri stars.  In {\refit Herbig-Haro Flows and
the Birth of Stars \/}, eds. \ B. Reipurth and C. Bertout (Kluwer
Academic Publishers), pp.\ 465-474.}

\ref{Guillot, T., Burrows, A., Hubbard, W. B., Lunine, J. I., and 
Saumon, D. 1996. Giant planets at small orbital distances.  {\refit
Astrophys.\ J.\/} 459:L35--L38.}

\ref{Hagihara, Y. 1972. {\refit Celestial Mechanics, Vol. II, 
Part 1: Perturbation theory \/} (Cambridge: MIT Press), p. \ 447.}

\ref{Hartmann, L., Calvet, N., Gullbring, E., and D'Alessio, P. 1998.
Accretion and the Evolution of T Tauri Disks.  {\refit Astrophys.\
J.\/} 495:385--400.}

\ref{Holman, M., Touma, J., and Tremaine, S. 1997. 
Chaotic variations in the eccentricity of the planet orbiting
16~Cyg~B.  {\refit Nature} 386:254.}

\ref{Hut, P. 1980. Stability of tidal equilibrium. 
{\refit Astron.\ Astrophys.\/} 92:167--170.}

\ref{Hut, P. 1981. Tidal evolution in close binary systems.
{\refit Astron.\ Astrophys.\/} 99:126--140.}

\ref{Ida, S., and Lin, D. N. C. 1998. {\refit Submitted}.}

\ref{Innanen, K. A., Zheng, J. Q., Mikkola, S., and Valtonen, 
M. J. 1997. The Kozai mechanism and the stability of planetary orbits
in binary star systems. {\refit Astron.\ J.\/} 113:1915--1919.}

\ref{Ivanov, P. B., Papaloizou, J. C. B., and Polnarev, A. G, 1998.
The evolution of supermassive binary caused by an accretion disc.
{\refit Mon.\ Not.\ Roy.\ Astron.\ Soc.\/} {\refit Submitted}}

\ref{Kley, W. 1998.
Mass flow and accretion through gaps in accretion disks.  {\refit
Mon.\ Not.\ Roy.\ Astron.\ Soc.\/} {\refit Submitted}.}

\ref{Kokubu, E., and Ida, S. 1996. 
On Runaway Growth of Planetesimals.  {\refit Icarus} 123:180--191.}

\ref{Konigl, A. 1991. Disk accretion onto magnetic T Tauri stars.
{\refit Astrophys.\ J.\/} 370:L39--L43.}

\ref{Korycansky, D. G., and  Papaloizou, J. C. B., 1996.
A Method for Calculations of Nonlinear Shear Flow: Application to
Formation of Giant Planets in the Solar Nebula.  {\refit Astrophys.\
J.\ Supp.\/} 105:181--190.}

\ref{Korycansky, D. G., and Pollack, J. B. 1993.
Numerical calculations of the linear response of a gaseous disk to a
protoplanet.  {\refit Icarus} 102:150--165.}

\ref{Kozai, Y. 1962.
Secular perturbations of asteroids with high inclination and
eccentricity.  {\refit Astron.\ J.\/} 67:591--598.}

\ref{Kumar, P., Ao, C. O., and Quataert, E. J. 1995.
Tidal Excitation of Modes in Binary Systems with Applications to
Binary Pulsars. {\refit Astrophys.\ J.\/} 449:294--309.}

\ref{Kumar, P., and Goodman, J. 1996. Nonlinear damping of 
oscillations in tidal-capture binaries. {\refit Astrophys.\ J.\/}
466:946--956.}

\ref{Lin, D. N. C., Bodenheimer, P., and  Richardson, D. C. 1996.
Orbital migration of the planetary companion of 51 Pegasi to its
present location. {\refit Nature} 380:606--607.}

\ref{Lin, D. N. C., and Ida, S. 1997.
On the Origin of Massive Eccentric Planets.  {\refit Astrophys.\ J.\/}
477:781--791.}

\ref{Lin, D. N. C., and  Papaloizou, J. C. B. 1979a.
Tidal torques on accretion discs in binary systems with extreme mass
ratios.  {\refit Mon.\ Not.\ Roy.\ Astron.\ Soc.\/} 186:799--496.}

\ref{Lin, D. N. C., and Papaloizou, J. C. B. 1979b. On the structure 
of circumbinary accretion disks and the tidal evolution of
commensurable satellites. {\refit Mon.\ Not.\ Roy.\ Astron.\ Soc.\/}
188:191--201.}

\ref{Lin, D. N. C., and  Papaloizou, J. C. B. 1980.
On the structure and evolution of the primordial solar nebula.
{\refit Mon.\ Not.\ Roy.\ Astron.\ Soc.\/} 191:37--48.}

\ref{Lin, D. N. C., and  Papaloizou, J. C. B. 1986a.
On the tidal interaction between protoplanets and the primordial
solar nebula. II - Self-consistent nonlinear interaction.
{\refit Astrophys.\ J.\/} 307:395--409.}

\ref{Lin, D. N. C., and  Papaloizou, J. C. B. 1986b.
On the tidal interaction between protoplanets and the primordial
solar nebula. III - Orbital migration of protoplanets.
{\refit Astrophys.\ J.\/} 309:846--857.}

\ref{Lin, D. N. C., and  Papaloizou, J. C. B. 1993.
On the tidal interaction between protostellar disks and companions.
In {\refit Protostars and Planets III \/}, eds. \ E. H. Levy and
J. I. Lunine (Tucson: University of Arizona Press), pp.\ 749--835.}

\ref{Lin, D. N. C., Papaloizou, J. C. B., and Ruden, S. P.  1987.
On the confinement of planetary arcs.  {\refit Mon.\ Not.\ Roy.\
Astron.\ Soc.\/} 227:75--95.}

\ref{Lin, D. N. C., Papaloizou, J. C. B., and Savonije, G. J.  1990a.
Wave propagation in gaseous accretion disks.  {\refit Astrophys.\
J.\/} 364:326--334.}

\ref{Lin, D. N. C., Papaloizou, J. C. B., and Savonije, G. J.  1990b.
Propagation of tidal disturbance in gaseous accretion disks.  {\refit
Astrophys.\ J.\/} 365:748--756.}

\ref{Lissauer, J. J., and Stewart, G. R. 1993.
Growth of planets from planetesimals.  In {\refit Protostars and
Planets III \/}, eds. \ E. H. Levy and J. I. Lunine (Tucson:
University of Arizona Press), pp.\ 1061--1088.}

\ref{Lubow, S. H., Tout, C. A., and Livio, M. 1997. 
Resonant tides in close orbiting planets. {\refit Astrophys.\ J.\/}
484:866--870.}

\ref{MacDonald, G. J. F. 1964. Tidal friction. 
{\refit Rev.\ Geophys.\/} 2:467--541.}

\ref{Marcy, G. W., Butler, R. P., Williams, E., Bildsten, L., Graham, 
J. R., Ghez, A. M., and Jernigan, J. G. 1997. The planet around 51 
Pegasi. {\refit Astrophys.\ J.\/} 481:926--935.}

\ref{Mathieu, R. D. 1994. Pre--main--sequence binary stars.
{\refit Ann.\ Rev.\ Astron.\ Astrophys.\/} 32:465--530.}

\ref{Mathieu et al. 1998.  {\refit Submitted.}

\ref{McCaughrean, M. J., and Stauffer, J. R. 1994.
High resolution near-infrared imaging of the trapezium: A stellar
census.  {\refit Astron.\ J.\/} 108:1382--1397.}

\ref{Miyoshi, K., et al. 1998.
Numerical simulation of tidal interaction between a protoplanet and
the solar nebula.  {\refit Submitted}.}

\ref{Mizuno, H. 1980.
Formation of the giant planets.  {\refit Prog.\ Theor.\ Phys.\/}
64:544--557.}

\ref{Papaloizou, J. C. B., and Lin, D. N. C. 1984.
On the tidal interaction between protoplanets and the primordial solar
nebula. I - Linear calculation of the role of angular momentum
exchange.  {\refit Astrophys.\ J.\/} 285:818--834.}

\ref{Papaloizou, J. C. B., and Savonije, G. J. 1985.
On the tidal evolution of massive X--ray binaries --- The tidal
evolution time--scales for very long orbital periods. {\refit Mon.\
Not.\ Roy.\ Astron.\ Soc.\/} 213:85--96.}

\ref{Papaloizou, J. C. B., and Savonije, G. J. 1997.  Non-adiabatic 
tidal forcing of a massive, uniformly rotating star ---
III. Asymptotic treatment for low frequencies in the inertial
regime. {\refit Mon.\ Not.\ Roy.\ Astron.\ Soc.\/} 291:651--657.}

\ref{Pollack, J. B., Hubickyj, O., Bodenheimer, P.,
Lissauer, J. J., Podolak, M., and Greenzweig, Y.  1996.  Formation of
the Giant Planets by Concurrent Accretion of Solids and Gas.  {\refit
Icarus} 124:62--85.}

\ref{Prosser, C. F., Stauffer, J. R., Hartmann, L., Soderblom, 
D. R., Jones, B. F., Werner, M. W., and McCaughrean, M. J. 1994.  HST
photometry of the trapezium cluster.  {\refit Astrophys.\ J.\/}
421:517--541.}

\ref{Rasio, F. A., and Ford, E. B. 1996.
Dynamical instabilities and the formation of extrasolar planetary
systems.  {\refit Science} 274:954--956.}

\ref{Rasio, F. A., Tout, C. A., Lubow, S. H., and Livio, M. 1996. 
Tidal decay of close planetary orbits.  {\refit Astrophys.\ J.\/}
470:1187--1191.}

\ref{Safronov, V. S.  1969.
In {\refit Evoliutsiia doplanetnogo oblaka. \/} (Moscow)}

\ref{Sandquist, E., Taam, R. E., Lin, D. N. C., and Burkert, A. 1998.
Planet Stripping and Stellar Metallicity Enhancements.  {\refit
Astrophys.\ J.\/} in press.}

\ref{Saumon, D., Hubbard, W. B., Burrows, A., Guillot, T., 
Lunine, J. I., and Chabrier, G. 1996. A Theory of Extrasolar Giant
Planets. {\refit Astrophys.\ J.\/} 460:993--1018.}

\ref{Savonije, G. J., and Papaloizou, J. C. B. 1983.
On the tidal spin up and orbital circularization rate for the massive
X-ray binary systems. {\refit Mon.\ Not.\ Roy.\ Astron.\ Soc.\/}
203:581--593.}

\ref{Savonije, G. J., and Papaloizou, J. C. B. 1984. 
On the tidal evolution of massive X--ray binaries --- The spin--up and
circularization rates for systems with evolved stars and the effects
of resonances.  {\refit Mon.\ Not.\ Roy.\ Astron.\ Soc.\/}
207:685--704.}

\ref{Savonije, G. J., and Papaloizou, J. C. B. 1997.
Non--adiabatic tidal forcing of a massive, uniformly rotating star ---
II. The low-frequency, inertial regime. {\refit Mon.\ Not.\ Roy.\
Astron.\ Soc.\/} 291:633--650.}

\ref{Savonije, G. J., Papaloizou, J. C. B., and Alberts, F. 1995.
Nonadiabatic tidal forcing of a massive uniformly rotating star.
{\refit Mon.\ Not.\ Roy.\ Astron.\ Soc.\/} 277:471--496.}

\ref{Shakura, N. I., and Sunyaev, R. A. 1973.
Black holes in binary systems: Observational appearance.
{\refit Astron.\  Astrophys.\/} 24:337--355.}

\ref{Shu, F. H. 1984.
Waves in planetary rings.  In {\refit Planetary Rings\/}, eds. \
R. Greenberg and A. Brahic (Tucson: University of Arizona Press), pp.\
513--561.}

\ref{Stevenson, D. J. 1983. Anomalous bulk viscosity of two-phase 
fluids and implications for planetary interiors. {\refit J.\ Geophys.\
Res.\/} 88:2445-2455.}

\ref{Takeuchi, T., Miyama, S. M., and  Lin, D. N. C. 1996.
Gap Formation in Protoplanetary Disks.  {\refit Astrophys.\ J.\/}
460:832--847.}

\ref{Tassoul, J.--L. 1988. On orbital circularization in detached 
close binaries. {\refit Astrophys.\ J.\/} 324:L71--L73.}

\ref{Terebey, S., Shu, F. H., and Cassen, P. 1984.
The collapse of the cores of slowly rotating isothermal clouds.
{\refit Astrophys.\ J.\/} 286:529--551.}

\ref{Terquem, C., Papaloizou, J. C. B., Nelson, R. P., and 
Lin, D. N. C. 1998a. On the Tidal Interaction of a Solar--Type Star
with an Orbiting Companion: Excitation of g--Mode Oscillation and
Orbital Evolution. {\refit Astrophys.\ J.\/} 502:788--801.}

\ref{Terquem, C., Papaloizou, J. C. B., Nelson, R. P., and 
Lin, D. N. C. 1998b. Oscillations in solar--type stars tidally induced
by orbiting planets. In {\refit Planetary Systems: The Long View
\/}, ed. \ L. Celnikier (Editions Fronti\`eres), {\refit in press}}

\ref{Ward, W. R. 1981. Solar nebula dispersal and the 
stability of the planetary system. I--- Scanning secular resonance
theory. {\refit Icarus} 47:234--264.}

\ref{Ward, W. R. 1993.
Disk-planet interactions: Torques from the coorbital zone.  {\refit
Anal.\ NY Acad.\ Sci.\/} 675:314--323.}

\ref{Ward, W. R. 1997.
Protoplanet Migration by Nebula Tides.  {\refit Icarus} 126:261--281.}

\ref{Weidenschilling, S. J., and Marzari, F. 1996.
Gravitational scattering as a possible origin for giant planets at
small stellar distances.  {\refit Nature} 384:619--621.}

\ref{Wetherill, G. W., and Stewart, G. R. 1989.
Accumulation of a swarm of small planetesimals.  {\refit Icarus}
77:330--357.}

\ref{Zahn, J. P. 1975.
The dynamical tide in close binaries.  {\refit Astron.\ Astrophys.\/}
41:329--344.}

\ref{Zahn, J. P. 1977.
Tidal friction in close binary stars. {\refit Astron.\ Astrophys.\/}
57:383--394.}

\ref{Zharkov, V. N., and Gudkova, T. V. 1991.
Models of giant planets with a variable ratio of ice to rock.  {\refit
Ann.\ Geophy.\/} 9:357--366.}

\bye

%% file: ppiv-style.tex
%
%
%
\font\ninerm=cmr9
\font\eightrm=cmr8
\font\sixrm=cmr6
\font\ninei=cmmi9
\font\eighti=cmmi8
\font\sixi=cmmi6
\skewchar\ninei='177 \skewchar\eighti='177 \skewchar\sixi='177
\font\ninesy=cmsy9
\font\eightsy=cmsy8
\font\sixsy=cmsy6
\skewchar\ninesy='60 \skewchar\eightsy='60 \skewchar\sixsy='60

\font\ninebf=cmbx9
\font\eightbf=cmbx8
\font\sixbf=cmbx6
\font\ninett=cmtt9
\font\eighttt=cmtt8
\hyphenchar\tentt=-1 
\hyphenchar\ninett=-1
\hyphenchar\eighttt=-1
\font\ninesl=cmsl9
\font\eightsl=cmsl8
\font\nineit=cmti9
\font\eightit=cmti8
\newskip\ttglue
\def\tenpoint{\def\rm{\fam0\tenrm}%
  \textfont0=\tenrm \scriptfont0=\sevenrm \scriptscriptfont0=\fiverm
  \textfont1=\teni \scriptfont1=\seveni \scriptscriptfont1=\fivei
  \textfont2=\tensy \scriptfont2=\sevensy \scriptscriptfont2=\fivesy
  \textfont3=\tenex \scriptfont3=\tenex \scriptscriptfont3=\tenex
  \def\it{\fam\itfam\tenit}%
  \textfont\itfam=\tenit
  \def\sl{\fam\slfam\tensl}%
  \textfont\slfam=\tensl
  \def\bf{\fam\bffam\tenbf}%
  \textfont\bffam=\tenbf \scriptfont\bffam=\sevenbf
   \scriptscriptfont\bffam=\fivebf
  \def\tt{\fam\ttfam\tentt}%
  \textfont\ttfam=\tentt
  \tt \ttglue=.5em plus.25em minus.15em
  \normalbaselineskip=12pt
  \let\sc=\eightrm
  \let\big=\tenbig
  \setbox\strutbox=\hbox{\vrule height8.5pt depth3.5pt width0pt}%
  \normalbaselines\rm}
\def\ninepoint{\def\rm{\fam0\ninerm}%
  \textfont0=\ninerm \scriptfont0=\sixrm \scriptscriptfont0=\fiverm
  \textfont1=\ninei \scriptfont1=\sixi \scriptscriptfont1=\fivei
  \textfont2=\ninesy \scriptfont2=\sixsy \scriptscriptfont2=\fivesy
  \textfont3=\tenex \scriptfont3=\tenex \scriptscriptfont3=\tenex
  \def\it{\fam\itfam\nineit}%
  \textfont\itfam=\nineit
  \def\sl{\fam\slfam\ninesl}%
  \textfont\slfam=\ninesl
  \def\bf{\fam\bffam\ninebf}%
  \textfont\bffam=\ninebf \scriptfont\bffam=\sixbf
   \scriptscriptfont\bffam=\fivebf
  \def\tt{\fam\ttfam\ninett}%
  \textfont\ttfam=\ninett
  \tt \ttglue=.5em plus.25em minus.15em
  \normalbaselineskip=10pt 
  \let\sc=\sevenrm
  \let\big=\ninebig
  \setbox\strutbox=\hbox{\vrule height8pt depth3pt width0pt}%
  \normalbaselines\rm}
\def\eightpoint{\def\rm{\fam0\eightrm}%
  \textfont0=\eightrm \scriptfont0=\sixrm \scriptscriptfont0=\fiverm
  \textfont1=\eighti \scriptfont1=\sixi \scriptscriptfont1=\fivei
  \textfont2=\eightsy \scriptfont2=\sixsy \scriptscriptfont2=\fivesy
  \textfont3=\tenex \scriptfont3=\tenex \scriptscriptfont3=\tenex
  \def\it{\fam\itfam\eightit}%
  \textfont\itfam=\eightit
  \def\sl{\fam\slfam\eightsl}%
  \textfont\slfam=\eightsl
  \def\bf{\fam\bffam\eightbf}%
  \textfont\bffam=\eightbf \scriptfont\bffam=\sixbf
   \scriptscriptfont\bffam=\fivebf
  \def\tt{\fam\ttfam\eighttt}%
  \textfont\ttfam=\eighttt
  \tt \ttglue=.5em plus.25em minus.15em
  \normalbaselineskip=9pt
  \let\sc=\sixrm
  \let\big=\eightbig
  \setbox\strutbox=\hbox{\vrule height7pt depth2pt width0pt}%
  \normalbaselines\rm}
%
\def\headtype{\ninepoint}                 
\def\abstracttype{\ninepoint}             
\def\captiontype{\ninepoint}              
\def\footnotetype{\ninepoint}             
\def\refit{\it}                           
\font\chaptitle=cmr10 at 11pt             
\rm                                       

%
%
\parindent=0.25in                         
\parskip=0pt                              
\baselineskip=12pt                        
\hsize=4.25truein                         
\vsize=7.445truein                        
\hoffset=1in                              
\voffset=-0.5in                           

\newskip\sectionskipamount                
\newskip\aftermainskipamount              
\newskip\subsecskipamount                 
\newskip\firstpageskipamount              
\newskip\capskipamount                    
\newskip\ackskipamount                    
\sectionskipamount=0.2in plus 0.09in
\aftermainskipamount=6pt plus 6pt         
\subsecskipamount=0.1in plus 0.04in
\firstpageskipamount=3pc
\capskipamount=0.1in
\ackskipamount=0.15in
\def\sectionskip{\vskip\sectionskipamount}
\def\aftermainskip{\vskip\aftermainskipamount}
\def\subsecskip{\vskip\subsecskipamount} 
\def\firstpageskip{\vskip\firstpageskipamount}
\def\capskip{\hskip\capskipamount}

%
%
\nopagenumbers                            
\newcount\firstpageno                     
\firstpageno=\pageno                      
\newcount\chapno                          

\def\rightheadline{\headtype\phantom{\folio}\hfil\runningtitletext\hfil\folio}
\def\leftheadline{\headtype\folio\hfil\runningauthortext\hfil\phantom{\folio}}
\headline={\ifnum\pageno=\firstpageno\hfil
           \else
              \ifdim\ht\topins=\vsize           
                 \ifdim\dp\topins=1sp \hfil     
                 \else
                     \ifodd\pageno\rightheadline\else\leftheadline\fi
                 \fi
              \else
                 \ifodd\pageno\rightheadline\else\leftheadline\fi
              \fi
           \fi}

\def\bottomnumber{\hss\tenrm[\folio]\hss}
\footline={\ifnum\pageno=\firstpageno\bottomnumber\else\hfil\fi}

%
%
%
%
\outer\def\mainsection#1
    {\vskip 0pt plus\smallskipamount\sectionskip
     \message{#1}\vbox{\noindent{\bf#1}}\nobreak\aftermainskip\noindent}
 
\outer\def\subsection#1
    {\vskip 0pt plus\smallskipamount\subsecskip
     \message{#1}\vbox{\noindent{\bf#1}}\nobreak\smallskip\nobreak\noindent}
 

\def\title#1{{\chaptitle\leftline{#1}}}
\def\name#1{\leftline{#1}}
\def\affiliation#1{\leftline{\it #1}}
\def\abstract#1{{\abstracttype \noindent #1 \smallskip\vskip .1in}}
\def\ref{\noindent \parshape2 0truein 4.25truein 0.25truein 4truein}
\def\caption{\noindent \captiontype
             \parshape=2 0truein 4.25truein .125truein 4.125truein}

\def\footnote#1{\edef\fspafac{\spacefactor\the\spacefactor}#1\fspafac
      \insert\footins\bgroup\footnotetype
      \interlinepenalty100 \let\par=\endgraf
        \leftskip=0pt \rightskip=0pt
        \splittopskip=10pt plus 1pt minus 1pt \floatingpenalty=20000
        \textindent{#1}\bgroup\strut\aftergroup\strut\egroup\let\next}
\skip\footins=12pt plus 2pt minus 4pt 
\dimen\footins=30pc 

%
%

\def\@{\spacefactor 1000}

\def\,{\pcomma} 
\def\pcomma{\relax\ifmmode\mskip\thinmuskip\else\thinspace\fi}

\def\oversim#1#2{\lower0.5ex\vbox{\baselineskip=0pt\lineskip=0.2ex
     \ialign{$\mathsurround=0pt #1\hfil##\hfil$\crcr#2\crcr\sim\crcr}}}